\documentclass[12pt]{article}

\usepackage[utf8]{inputenc}
\usepackage[T1]{fontenc}
\usepackage{amsmath,amssymb}
\usepackage{bm}
\usepackage{url}
\usepackage{geometry}
\usepackage{natbib}
\usepackage{authblk}
\usepackage{amsthm}
\usepackage{graphicx}

\theoremstyle{definition}        
\newtheorem{theorem}{Theorem}
\title{Smooth SCAD:
A Raised Cosine SCAD Type Thresholding Rule for Wavelet Denoising}
\author{
Radhika Kulkarni$^{1}$,
Aluisio Pinheiro$^{2}$,
Brani Vidakovic$^{1}$, and   Abdourrahmane M. Atto$^{3}$ \\
 
{\small $^{1}$Department of Statistics, Texas A\&M University, College Station, TX 77843, USA} \\

{\small $^{2}$Departamento de Estat\'istica, UNICAMP, Campinas 13083-859, Brasil}
\\

{\small $^{3}$Universit\'e Savoie Mont Blanc, Polytech Annecy-Chamb\'ery, 74944 Annecy Cedex, France }
}

\date{\today}

\begin{document}
\maketitle

\begin{abstract}
We introduce a smooth variant of the SCAD thresholding rule for wavelet denoising by replacing its piecewise linear transition with a raised cosine. The resulting shrinkage function is odd, continuous on $\mathbb{R}$, and continuously differentiable away from the main threshold, yet retains the hallmark SCAD properties of sparsity for small coefficients and near unbiasedness for large ones. This smoothness places the rule within the continuous thresholding class for which Stein’s unbiased risk estimate is valid. As a result, unbiased risk computation, stable data-driven threshold selection, and the asymptotic theory of Kudryavtsev and Shestakov apply.
 A corresponding nonconvex prior is obtained whose posterior mode coincides with the estimator, yielding a transparent Bayesian interpretation. We give an explicit SURE risk expression, discuss the oracle scale of the optimal threshold, and describe both global and level-dependent adaptive versions. The smooth SCAD rule therefore offers a tractable refinement of SCAD, combining low bias, exact sparsity, and analytical convenience in a single wavelet shrinkage procedure.

\medskip

\end{abstract}


\section{Introduction}

Wavelet thresholding is a central tool for nonparametric noise removal and signal recovery. After an orthogonal wavelet transform, the empirical coefficients $d_{j,k}$ follow the Gaussian sequence model
\begin{eqnarray}
d_{j,k} \sim N(\theta_{j,k},\sigma^{2}),
\label{eq:basicmodel}
\end{eqnarray}
and the task is to estimate the unknown coefficients $\theta_{j,k}$ while keeping the estimator stable and essentially unbiased for genuinely large coefficients. Since the work of Donoho and Johnstone, a wide spectrum of wavelet thresholding rules has been studied, ranging from simple hard and soft rules to nonconvex rules motivated by penalized likelihood and Bayesian ideas; see for example \citet{AntoniadisFan2001,AntoniadisSapatinas2007,AntoniadisBigotSapatinas2007}.

A particularly influential contribution is the SCAD penalty of \citet{FanLi2001}, introduced for variable selection with nonconcave penalized likelihood. SCAD was designed to combine three properties: sparsity through an exact zero region, continuity of the estimator, and near unbiasedness for large coefficients. These features make the associated threshold very natural for wavelet shrinkage, as emphasized by \citet{AntoniadisFan2001}. The SCAD threshold behaves like soft thresholding near the main threshold and like the identity for large coefficients, creating a practical compromise between bias and variance. This construction has inspired related nonconvex penalties, including the minimax concave penalty \citep{Zhang2010}, one--step sparse estimators \citep{ZouLi2008}, and coordinate--descent based procedures such as SparseNet and its variants \citep{Mazumder2011,BrehenyHuang2011,HuangBrehenyMa2013}. SCAD and other thresholding methods form a relevant issue in the context of wavelet feature screening for high and ultra-high dimensional data \citep{Fonseca2024} as well.

In the wavelet and inverse problems domain, continuous and nonconvex thresholding rules have been studied in a variety of settings. Examples include wavelet regularization and iterative thresholding algorithms for linear inverse problems \citep{DaubechiesDefriseDeMol2004}, robust wavelet shrinkage and Bayesian refinements \citep{SardyTsengBruce2001,SardyBruce2003}, and nonconvex penalized estimators for denoising and regression \citep{ZhouShen2008,LiPengZhang2011}. Smoothly clipped and smoothed $\ell_{0}$--type penalties have also been proposed to obtain continuous shrinkage rules with good sparsity properties \citep{Xue2009,GassoRakotomamonjy2011}. These contributions highlight the benefits of continuity and smoothness in the shrinkage function for both numerical stability and statistical performance.

From this perspective the classical SCAD rule fits well in applications, but has a small analytical limitation. The resulting shrinkage function is continuous but not continuously differentiable at its internal breakpoints. This causes no numerical difficulty; however, it prevents the rule from fitting within the continuous SURE framework developed by \citet{KudryavtsevShestakov2016,Shestakov2020,KudryavtsevShestakov2024}, which we collectively refer to as the KS framework. 
Their theory provides unbiased risk estimation (via Stein identities) for a broad class of
continuous thresholding rules written in the form
$$
\rho_h(d,\lambda)
=
\left\{
\begin{array}{ll}
d-h(d,\lambda), & |d|>\lambda,\\[0.2em]
0, & |d|\le \lambda,
\end{array}
\right.
$$
where the auxiliary function $h(d,\lambda)$ satisfies three structural requirements:
(i) oddness $h(-d,\lambda)=-h(d,\lambda)$; (ii) boundedness $0\le h(d,\lambda)\le \lambda$ for $d\ge0$;
and (iii) threshold matching $h(\lambda,\lambda)=\lambda$.
In addition, KS impose uniform derivative bounds  of the form
$|\partial h/\partial d|\le C\lambda^\mu$, $|\partial h/\partial\lambda|\le C\lambda^\mu$, and
$|\partial^2 h/(\partial d\,\partial\lambda)|\le C\lambda^\mu$,
which ensure the asymptotic theory for SURE-selected thresholds.
 Because the SCAD shrinkage profile has kinks, it formally falls just outside the assumptions required for their asymptotic analysis; while other frameworks such as the sigmoid shrinkage \citep{atto2008sigmoid} have led to a safe SURE risk estimation and the use of this risk to derive optimal sigmoid parameters  \citep{atto2009sure}.

The goal of this note is to place the spirit of SCAD firmly inside that continuous SURE admissible class. We introduce a smooth variant by replacing the linear transition between $\lambda$ and $a\lambda$ with a raised cosine. 
This produces a shrinkage amount $h(d,\lambda)$ that is odd and uniformly bounded, and that is continuous and continuously differentiable throughout the transition region $\lambda<|d|<a\lambda$ (with continuity at the junction points $|d|=\lambda$ and $|d|=a\lambda$ as required by the KS framework).

The resulting estimator behaves like SCAD in the small and large coefficient regimes, but is smooth in between.
Because the cosine transition has bounded derivatives on $\lambda<|d|<a\lambda$, and $h$ is constant/zero outside that region, the rule satisfies the structural conditions (oddness, boundedness, and continuity at $|d|=\lambda$) and the derivative bounds required in \citet{KudryavtsevShestakov2024}. In particular, the smooth SCAD rule belongs to their admissible class of continuous thresholding functions for which risk bounds, asymptotic normality of the SURE risk, and strong consistency are available.

The wavelet domain has long been a natural setting for continuous and semisoft thresholding schemes. Robust and adaptive methods \citep{SardyTsengBruce2001,SardyBruce2003}, nonconvex penalties \citep{ZhouShen2008,LiPengZhang2011}, and empirical Bayes constructions \citep{FigueiredoNowak2003} all exploit sparsity or near--sparsity of wavelet coefficients. Smooth SCAD follows the same philosophy. It keeps the desirable behavior of SCAD, namely sparsity, continuity, and near unbiasedness for large coefficients, but eliminates abrupt slope changes that complicate analytical treatment.

\medskip
\noindent\textbf{Explicit contributions.}
While SCAD has been repeatedly advocated for wavelet shrinkage, no continuously differentiable SCAD--type rule has previously been shown to satisfy the full admissibility conditions of the continuous SURE framework. The present paper fills this gap. The main contributions are:
\begin{itemize}
\item[(i)] A raised cosine smooth SCAD generator $h(d,\lambda)$ is constructed and shown to satisfy the structural conditions of the KS framework, together with uniform derivative bounds, placing a SCAD-type rule (with a raised-cosine transition) inside the continuous SURE class.

\item[(ii)] An explicit SURE formula specialized to smooth SCAD is derived, enabling exact evaluation of the unbiased risk for any fixed threshold.
\item[(iii)] A Bayesian interpretation is provided, based on a smooth nonconvex penalty whose posterior mode reproduces the estimator exactly.
\item[(iv)] Threshold selection methodology is developed: global SURE selection, oracle asymptotics at the scale $\sqrt{\log N}$, simple analytic approximations, and 
level-dependent adaptive versions, with smooth SCAD on the same SURE–based footing as benchmark rules such as SureShrink.
\end{itemize}

On the Bayesian side, we identify an explicit smooth SCAD prior with effectively flat tails and a closed form penalty whose posterior mode coincides with the raised cosine threshold and reveals the exact point $|d|=\lambda$ at which the nonzero mode is born.

\medskip
The paper is organized as follows. 
Section~\ref{sec:model} introduces the wavelet model and the SURE framework. Section~\ref{sec:smoothscad-definition} defines smooth SCAD and gives the raised cosine construction. Section~\ref{sec:verification} verifies the structural assumptions and derivative bounds needed for SURE admissibility. Section~\ref{sec:risk-sure-asymptotics} derives the specialized SURE expression and summarizes the resulting asymptotic behavior. Section~\ref{sec:bayes} presents the Bayesian interpretation and induced prior. Section~\ref{sec:lambda-choice} discusses threshold and parameter selection, including oracle asymptotics, SURE--driven tuning, explicit analytic formulas, and level-dependent extensions. Section~\ref{sec:numerical} outlines a numerical illustration. Section~\ref{sec:conclusions} concludes.


\section{Model and SURE framework}
\label{sec:model}

We work directly in the wavelet domain. After an orthogonal discrete wavelet transform of a noisy signal of length $N$, the empirical wavelet coefficients satisfy
\begin{eqnarray}
d_{j,k} = \theta_{j,k}+\epsilon_{j,k}, ~~~\epsilon_{j,k} \sim N(0, \sigma^2),
\label{eq:normalmodel}
\end{eqnarray}
with the $\epsilon_{j,k}$ independent. The pairs $(j,k)$ index resolution levels $j=J-L,\ldots,J-1$ and locations $k=0,\ldots,2^{j}-1$, and $\theta_{j,k}$ are the unknown true coefficients. Here,
 $N=2^J$ and $L$ denotes the depth of wavelet transform.
Thresholding is applied coefficientwise on $L$ detail multiresolution levels using a function $\rho(d,\lambda)$ with a global threshold $\lambda>0$. According to the KS framework, it is convenient to define $\rho$ through an auxiliary function $h(d,\lambda)$ satisfying three structural properties. First, $h$ is odd,
\begin{eqnarray}
h(-d,\lambda) &=& - h(d,\lambda),
\label{eq:odd}
\end{eqnarray}
second, for $d \ge 0$ it is bounded,
\begin{eqnarray}
0 \le h(d,\lambda) \le \lambda,
\label{eq:bound}
\end{eqnarray}
and third, it satisfies the threshold matching condition,
\begin{eqnarray}
h(\lambda,\lambda) &=& \lambda.
\label{eq:cont}
\end{eqnarray}
Given $h$, shrinkage is implemented through
\begin{eqnarray}
\rho_{h}(d,\lambda) &=&
\begin{cases}
0, & |d| \le \lambda, \\
d - h(d,\lambda), & |d| > \lambda,
\end{cases}
\label{eq:rho-def}
\end{eqnarray}
a formulation that includes soft thresholding, tanh and sigmoid rules, hybrid rules, and nonconvex SCAD--type procedures.

The mean squared error (risk) of wavelet thresholding at resolution depth $L$ is
\begin{eqnarray}
R_{J}(\lambda) &=& 
\sum_{j=J-L}^{J-1} \sum_{k=0}^{2^{j}-1}
\mathbb{E}\Bigl(\rho_{h}(d_{j,k},\lambda) - \theta_{j,k}\Bigr)^{2}.
\label{eq:risk}
\end{eqnarray}
Because $h$ will be differentiable in $d$ for $|d|>\lambda$ in our construction, Stein's identity yields an unbiased estimator of \eqref{eq:risk}. Define
\begin{eqnarray}
g(d,\lambda) &=& d - \rho_{h}(d,\lambda),
\label{eq:g}
\end{eqnarray}
which satisfies
\begin{eqnarray}
g(d,\lambda) &=&
\begin{cases}
d, & |d| \le \lambda, \\
h(d,\lambda), & |d| > \lambda.
\end{cases}
\label{eq:g-piece}
\end{eqnarray}
The SURE estimator is
\begin{eqnarray}
\widehat{R}_{J}(\lambda) &=&
\sum_{j=J-L}^{J-1} \sum_{k=0}^{2^{j}-1} F(d_{j,k},\lambda),
\label{eq:riskhat}
\end{eqnarray}
where
\begin{eqnarray}
F(d,\lambda) &=&
\begin{cases}
d^{2} - \sigma^{2}, & |d|\le \lambda, \\
h(d,\lambda)^{2} 
+ \sigma^{2}
- 2\sigma^{2}\dfrac{\partial h}{\partial d}(d,\lambda), & |d|>\lambda.
\end{cases}
\label{eq:F}
\end{eqnarray}
The SURE selected threshold is defined by
\begin{eqnarray}
\lambda_{S} &=&
\arg\min_{\lambda \in [0,\lambda_{U}]}\, \widehat{R}_{J}(\lambda),
\label{eq:lambdaS}
\end{eqnarray}
where $\lambda_{U}$ denotes the universal upper bound of order $\sigma\sqrt{2\log N}$.

The KS framework provides risk bounds, asymptotic normality, and consistency for $\lambda_{S}$ and $\widehat{R}_{J}(\lambda_{S})$ whenever $h$ satisfies \eqref{eq:odd}–\eqref{eq:cont} and mild regularity conditions on its derivatives. In Section~\ref{sec:smoothscad-definition} we introduce a smooth SCAD thresholding rule and verify in Section~\ref{sec:verification} that it falls inside this admissible class.

\section{Smooth SCAD via a raised cosine}
\label{sec:smoothscad-definition}

We construct a shrinkage rule with the qualitative behavior of SCAD, but with a smooth transition in place of its piecewise linear segments. We introduce a lower threshold $\lambda>0$ and an upper multiplicative factor $a>1$ that controls the width of the transition zone. Throughout, $a>1$ is treated as fixed.

Smooth SCAD is defined through the auxiliary function $h(d,\lambda)$, which generates the thresholding rule $\rho_{h}$ via \eqref{eq:rho-def}. We first specify $h$ for $d\ge 0$ and then extend it to $\mathbb{R}$ by oddness. For $d\ge 0$ we set
\begin{eqnarray}
h(d,\lambda) &=&
\begin{cases}
\lambda, & 0 < d \le \lambda, \\[0.2cm]
\lambda \cos^{2}\!\left(\dfrac{\pi}{2}\,\dfrac{d-\lambda}{(a-1)\lambda}\right), 
& \lambda < d < a\lambda, \\[0.3cm]
0, & d \ge a\lambda,
\end{cases}
\label{eq:h-smoothscad-pos}
\end{eqnarray}
with $h(0,\lambda)=0$. Supplementing the definition of $h$ with
\begin{eqnarray}
h(-d,\lambda) &=& -\,h(d,\lambda), \qquad d\ge 0,
\label{eq:h-smoothscad-odd}
\end{eqnarray}
guarantees $h$ is odd on $\mathbb{R}$. The branch $h(d,\lambda)=\lambda$ for $0<d\le\lambda$ is chosen to satisfy $h(\lambda,\lambda)=\lambda$ and to simplify the induced penalty integral; it does not affect the thresholding rule itself, since $\rho_{h}(d,\lambda)=0$ for $|d|\le\lambda$. In particular, any discontinuity of $h$ at $d=0$ is immaterial for the estimator because $\rho_h(d,\lambda)=0$ for $|d|\le\lambda$, and only the behavior of $h$ for $|d|>\lambda$  enters the KS regularity analysis.

To interpret \eqref{eq:h-smoothscad-pos}, define the normalized transition parameter
\begin{eqnarray*}
s(d, \lambda) &=& \frac{d-\lambda}{(a-1)\lambda}, \qquad \lambda < d < a\lambda,
\label{eq:s-param}
\end{eqnarray*}
so that $\lambda<d<a\lambda$ corresponds to $0<s<1$. Then
\begin{eqnarray*}
h(d,\lambda) &=& \lambda \cos^{2}\!\left(\frac{\pi}{2}s(d,\lambda)\right),
\qquad \lambda < d < a\lambda,
\label{eq:h-cos}
\end{eqnarray*}
which decreases smoothly from $\lambda$ at $d=\lambda$ to $0$ at $d=a\lambda$.

Given $h$, the smooth SCAD thresholding function is
\begin{eqnarray}
\rho_{\text{ssc}}(d,\lambda) &=&
\begin{cases}
0, & |d| \le \lambda,\\[0.1cm]
d - h(d,\lambda), & \lambda < |d| < a\lambda,\\[0.1cm]
d, & |d| \ge a\lambda.
\end{cases}
\label{eq:rho-ssc}
\end{eqnarray}
Thus $\rho_{\text{ssc}}$ is:
(i) exactly zero below the main threshold $\lambda$,  
(ii) smoothly interpolating in the region $\lambda < |d| < a\lambda$, and  
(iii) identical to the identity mapping when $|d| \ge a\lambda$.  

If the cosine transition in \eqref{eq:h-smoothscad-pos} were replaced by a linear ramp from $\lambda$ to $0$ on $[\lambda,a\lambda]$, the resulting $\rho_{h}$ would reproduce the classical SCAD rule of \citet{FanLi2001}. Smooth SCAD may therefore be viewed as a differentiable analogue of SCAD with the same small and large coefficient behavior but a smooth transition in between.

\begin{figure}[t]
\centering

\begin{minipage}{0.45\textwidth}
\centering
\includegraphics[width=\linewidth]{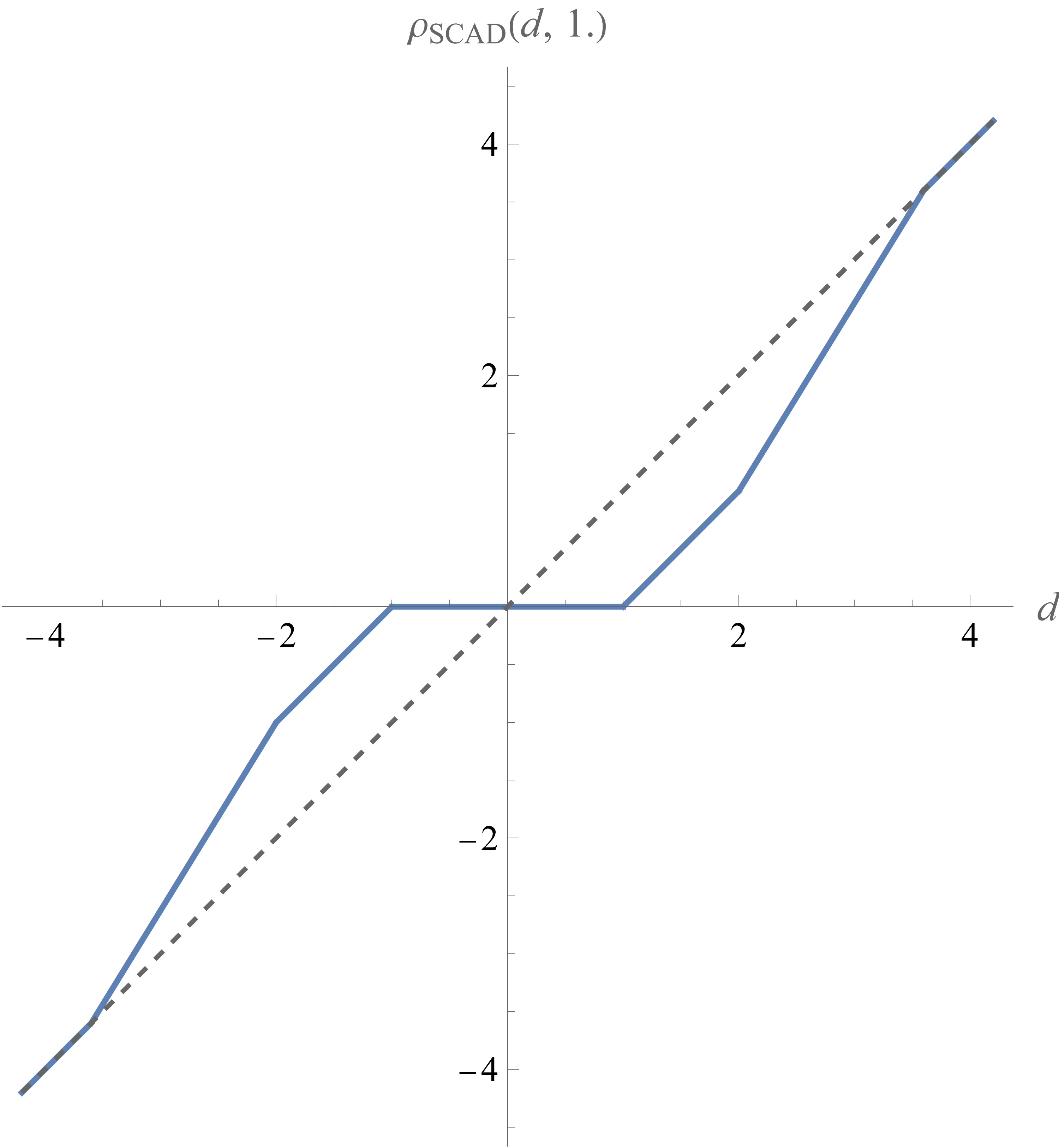}
\\[-2mm]
{\small (a)}
\end{minipage}
\hfill
\begin{minipage}{0.45\textwidth}
\centering
\includegraphics[width=\linewidth]{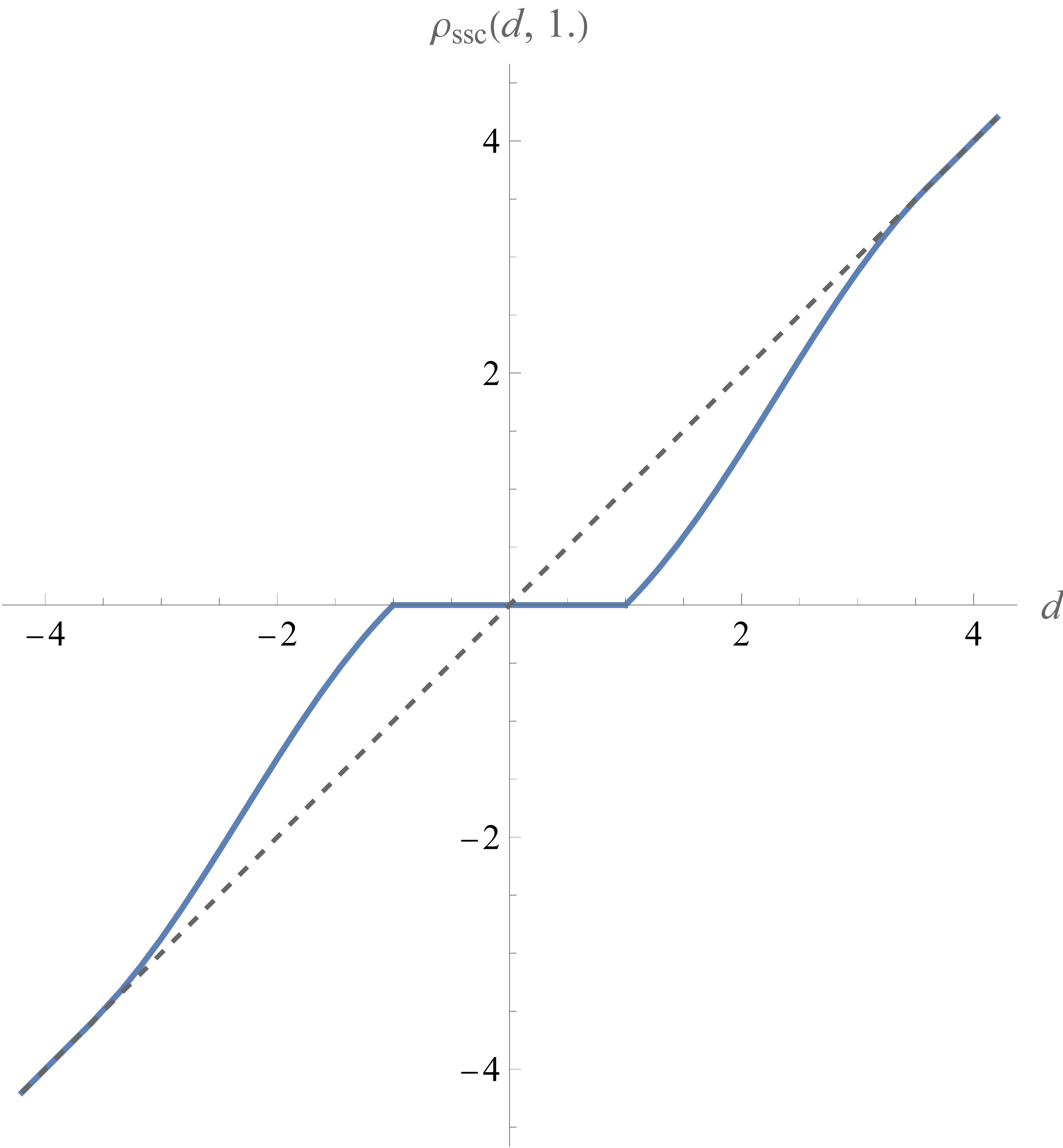}
\\[-2mm]
{\small (b)}
\end{minipage}

\caption{Classical SCAD (a) and smooth SCAD (b) thresholding functions for $\lambda=1$.}
\label{fig:scad-smoothscad}
\end{figure}

The function $\rho_{\text{ssc}}$ inherits oddness and continuity from \eqref{eq:h-smoothscad-pos}–\eqref{eq:h-smoothscad-odd}. 
In particular, $\rho_{\mathrm{ssc}}$ is continuous on $\mathbb{R}$, continuously differentiable for $|d|\notin\{\lambda,a\lambda\}$, and satisfies $\rho_{\mathrm{ssc}}(d,\lambda)=d$ for $|d|\ge a\lambda$, so it is exactly unbiased in the large-coefficient regime.
 The next step is to verify that $h$ satisfies the assumptions \eqref{eq:odd}–\eqref{eq:cont} and the differentiability and boundedness conditions required by the SURE theory in the KS framework. We summarize these properties in the next section.


\section{Verification of structural assumptions and regularity conditions}
\label{sec:verification}

We verify that the smooth SCAD construction defined in
\eqref{eq:h-smoothscad-pos}–\eqref{eq:rho-ssc} satisfies the assumptions required by the continuous SURE framework. All notation and definitions are as introduced in Sections~\ref{sec:model} and \ref{sec:smoothscad-definition}.

\subsection{Structural conditions}

The admissible SURE class is defined by three requirements on the shrinkage generator $h$: oddness, boundedness on $[0,\infty)$, and continuity at the main threshold $d=\lambda$ (conditions \eqref{eq:odd}–\eqref{eq:cont}); global continuity of $h$ is not required, since $\rho_h(d,\lambda)=0$ for $|d|\le\lambda$ and the derivative conditions are imposed only on $|d|>\lambda$.

Oddness follows immediately from the extension rule \eqref{eq:h-smoothscad-odd}. For $d\ge 0$, the cosine term in \eqref{eq:h-smoothscad-pos} satisfies $0\le \cos^{2}(\cdot)\le 1$, hence
\begin{eqnarray*}
0 \le h(d,\lambda) \le \lambda, \qquad d\ge 0,
\end{eqnarray*}
so boundedness holds.

Continuity at $d=\lambda$ is verified by
\begin{eqnarray*}
\lim_{d\downarrow \lambda} h(d,\lambda) &=& \lambda\cos^{2}(0) = \lambda, \\
\lim_{0<d\uparrow \lambda} h(d,\lambda) &=& \lambda,
\end{eqnarray*}
hence $h(\lambda,\lambda)=\lambda$ and continuity at the threshold holds. Continuity at $d=a\lambda$ is immediate because $h(a\lambda,\lambda)=0$ from the cosine branch and $h(d,\lambda)=0$ for $d\ge a\lambda$.  

Therefore smooth SCAD satisfies the structural conditions of the SURE theory.

\subsection{Derivative bounds}

The KS framework imposes boundedness conditions on derivatives of $h$ in both $d$ and $\lambda$, namely
\begin{eqnarray}
\left|\frac{\partial h}{\partial d}(d,\lambda)\right| \le C \lambda^{\mu}, \quad
\left|\frac{\partial h}{\partial \lambda}(d,\lambda)\right| \le C \lambda^{\mu}, \quad
\left|\frac{\partial^{2} h}{\partial d \partial \lambda}(d,\lambda)\right| \le C \lambda^{\mu},
\label{eq:derivbounds-sc}
\end{eqnarray}
for some constants $C>0$ and $\mu\ge 0$. It is enough to evaluate these derivatives for $d>0$; symmetry through \eqref{eq:h-smoothscad-odd} yields the corresponding bounds for $d<0$. 
Derivatives at the junction points $d=\pm\lambda,\pm a\lambda$ are not required by the KS framework, which imposes the derivative bounds only on the open region $|d|>\lambda$.

On the regions $0<|d|\le \lambda$ and $|d|\ge a\lambda$, $h$ is either constant or zero, hence
\begin{eqnarray*}
\frac{\partial h}{\partial d} = \frac{\partial h}{\partial \lambda}
= \frac{\partial^{2} h}{\partial d \partial \lambda} = 0.
\end{eqnarray*}

In the transition region $\lambda<d<a\lambda$,
\begin{eqnarray*}
h(d,\lambda) &=& \lambda \cos^{2}\!\left(\alpha\,s(d,\lambda)\right), \qquad
\alpha = \frac{\pi}{2}, \qquad s(d,\lambda)=\frac{d-\lambda}{(a-1)\lambda}.
\end{eqnarray*}
Differentiating with respect to $d$ yields
\begin{eqnarray*}
\frac{\partial h}{\partial d}(d,\lambda)
&=& -\frac{\alpha}{a-1} \sin\!\bigl(2\alpha\,s(d,\lambda)\bigr),
\end{eqnarray*}
hence
\begin{eqnarray*}
\left|\frac{\partial h}{\partial d}(d,\lambda)\right| \le \frac{\alpha}{a-1}
= \frac{\pi}{2(a-1)},
\end{eqnarray*}
which is uniform in $d$ and $\lambda$. Therefore the derivative bounds required by the KS conditions hold with exponent $\mu=0$.

For $\partial h/\partial\lambda$, differentiation of
$
h(d,\lambda)=\lambda\cos^2(\alpha s(d,\lambda))
$
with $s(d,\lambda)=(d-\lambda)/((a-1)\lambda)$ yields a finite sum of terms of the form
\[
\cos(\alpha s),\quad \sin(\alpha s),\quad
\frac{d-\lambda}{\lambda},\quad \frac{d}{\lambda},
\]
multiplied by constants depending only on $a$. On the transition region
$\lambda<d<a\lambda$ one has
\[
1<\frac{d}{\lambda}<a,
\qquad
0<\frac{d-\lambda}{(a-1)\lambda}<1,
\]
so all such terms are uniformly bounded independently of $d$ and $\lambda$. Hence
\[
\left|\frac{\partial h}{\partial\lambda}(d,\lambda)\right|\le C(a).
\]
The mixed derivative $\partial^2 h/(\partial d\,\partial\lambda)$ involves the same
bounded factors and therefore satisfies
\[
\left|\frac{\partial^2 h}{\partial d\,\partial\lambda}(d,\lambda)\right|\le C(a).
\]
Consequently, the regularity conditions \emph{(16)} hold with exponent $\mu=0$.

\medskip

In conclusion, the smooth SCAD rule belongs to the admissible continuous thresholding class in the KS framework. Its SURE risk, its SURE selected threshold, and its asymptotic behavior are governed by the general theory developed for that framework.


\section{Risk, SURE, and asymptotic behavior of smooth SCAD}
\label{sec:risk-sure-asymptotics}

For the smooth SCAD rule, the oracle risk is defined as in \eqref{eq:risk} with $\rho_{h}(d,\lambda)=\rho_{\text{ssc}}(d,\lambda)$ from \eqref{eq:rho-ssc}. Because $R_{J}(\lambda)$ depends on the unknown coefficients $\theta_{j,k}$ it is not available in closed form, but its unbiased estimator follows directly from \eqref{eq:riskhat}–\eqref{eq:F}. Substituting $h$ from \eqref{eq:h-smoothscad-pos}–\eqref{eq:h-smoothscad-odd} into the general expression gives
\begin{eqnarray}
\widehat{R}_{J}(\lambda)
&=& \sum_{j=J-L}^{J-1} \sum_{k=0}^{2^{j}-1} 
F_{\text{ssc}}(d_{j,k},\lambda),
\label{eq:suresc}
\end{eqnarray}
where
\begin{eqnarray}
F_{\text{ssc}}(d,\lambda)
&=&
\begin{cases}
d^{2} - \sigma^{2}, & |d|\le \lambda,\\[0.2cm]
h(d,\lambda)^{2} + \sigma^{2} - 2\sigma^{2}\dfrac{\partial h}{\partial d}(d,\lambda),
& \lambda < |d| < a\lambda,\\[0.3cm]
\sigma^{2}, & |d| \ge a\lambda,
\end{cases}
\label{eq:Fssc}
\end{eqnarray}
and $h$ and $\partial h/\partial d$ on the intermediate region are explicit:
\begin{eqnarray*}
h(d,\lambda) &=& \lambda \cos^{2}\!\left(\frac{\pi}{2}\,\frac{d-\lambda}{(a-1)\lambda}\right), 
\qquad \lambda<d<a\lambda,\\
\frac{\partial h}{\partial d}(d,\lambda)
&=& -\frac{\pi}{2(a-1)}
\sin\!\left(\pi \frac{d-\lambda}{(a-1)\lambda}\right),
\label{eq:dh-ssc}
\end{eqnarray*}
and for negative $d$ the odd extension \eqref{eq:h-smoothscad-odd} applies. Thus \eqref{eq:suresc} can be evaluated numerically without approximation for any realization of $(d_{j,k})$.

The SURE selected threshold is
\begin{eqnarray}
\lambda_{S} &=& \arg\min_{\lambda\in[0,\lambda_{U}]}\widehat{R}_{J}(\lambda),
\label{eq:lambdas-ssc}
\end{eqnarray}
with $\lambda_{U}$ the usual universal upper bound of order $\sigma\sqrt{2\log N}$. In practice, one evaluates $\widehat{R}_{J}(\lambda)$ on a grid and chooses the minimizing $\lambda$.

Because smooth SCAD satisfies the structural requirements \eqref{eq:odd}–\eqref{eq:cont} and the derivative bounds \eqref{eq:derivbounds-sc} with $\mu=0$, the asymptotic results of \citet{KudryavtsevShestakov2016,Shestakov2020,KudryavtsevShestakov2024} apply. In particular:

\begin{theorem}[Asymptotic behavior under Lipschitz regularity]
\label{thm:asymptotics}
Assume the underlying function $f$ is supported on a finite interval and uniformly Lipschitz with exponent $\gamma>1/2$. Then its true wavelet coefficients satisfy a sparsity--inducing decay condition of the usual form. For smooth SCAD shrinkage with fixed $a>1$, the oracle optimal risk $R_{J}(\lambda_{\min})$ obeys an upper bound of the same order as in the general continuous SURE class; in particular, there exists a constant $C$ depending on $\gamma$ and $a$ such that
\begin{eqnarray}
R_{J}(\lambda_{\min})
&\le&
C\,2^{J/(2\gamma+1)}\,J^{(2\gamma+2)/(2\gamma+1)}
\label{eq:riskrate}.
\end{eqnarray}
The precise constants and logarithmic factors are inherited directly from the KS analysis and depend only on $\gamma$ and $a$.

Moreover, the SURE selected threshold $\lambda_{S}$ avoids suboptimally small values with probability approaching one as $J\to\infty$, $R_{J}(\lambda_{S})$ is of the same order as $R_{J}(\lambda_{\min})$, and the centered SURE risk admits an asymptotically normal limit and a strong consistency property   \citep{KudryavtsevShestakov2024}.
\end{theorem}

The theorem summarizes what smooth SCAD inherits from the continuous SURE admissible class. 


\section{Bayesian interpretation of smooth SCAD}
\label{sec:bayes}

Bayesian methodology offers a natural and coherent framework for wavelet domain inference. Within this approach, Bayes rules act as shrinkage operators: prior specification governs the degree of coefficient attenuation, and loss functions determine the functional form of the rule. Under quadratic loss, posterior means give smooth shrinkage; alternative loss structures yield nonlinear shrinkers of thresholding type, often realized as posterior modes or medians. Thus Bayesian shrinkage encompasses both continuous and selective coefficient suppression within a unified paradigm, guided by explicit probabilistic modeling of signal and noise.

The Bayesian wavelet literature follows this line. \citet{Vidakovic1994} introduced Bayes rules and Bayes factors for adaptive shrinkage in wavelet settings. An explicit rule exhibiting threshold--like behavior through closed--form posterior modes was later developed   \citep{CutilloJungRuggeriVidakovic2008,VidakovicRuggeri2001_BAMS}, and its extensions reviewed in \citet{RemenyiVidakovic2013}. A broader foundation for model--based wavelet analysis is provided in \citet{Vidakovic1999}, which positions wavelet shrinkage within a multiscale inferential framework. Collectively, these works establish Bayesian shrinkage as a principled avenue for noise reduction, smoothing, and multiscale structure extraction.

We now place smooth SCAD into a simple Bayesian framework. For clarity, we work with a single coefficient and then indicate how this extends to the full wavelet vector.

Consider the normal means model
\begin{eqnarray}
d \mid \theta &\sim& N(\theta,\sigma^{2}),
\label{eq:bayes-lik}
\end{eqnarray}
and recall that the smooth SCAD estimator is
\begin{eqnarray}
\delta_{\text{ssc}}(d,\lambda) &=& \rho_{\text{ssc}}(d,\lambda),
\end{eqnarray}
where the structure of $\rho_{\text{ssc}}$ is given in \eqref{eq:rho-ssc} and depends on $h$ from \eqref{eq:h-smoothscad-pos}–\eqref{eq:h-smoothscad-odd}. The goal is to construct a prior $p(\theta)$ such that the posterior mode satisfies
\[
\hat{\theta}(d,\lambda) = \rho_{\text{ssc}}(d,\lambda).
\]

\subsection{Prior induced by the shrinkage profile}

The connection between smooth SCAD and Bayesian MAP estimation is obtained by integrating the shrinkage profile $h$. Define the penalty function on $r=|\theta|$ by
\begin{eqnarray}
\Phi(r;\lambda) &=& \int_{0}^{r} h(u,\lambda)\,du, \qquad r \ge 0,
\label{eq:Phi-def}
\end{eqnarray}
which is finite for all $r$ because $h$ is bounded and supported on $[0,a\lambda]$. The induced prior is
\begin{eqnarray}
p(\theta \mid \lambda,\sigma^{2}) &\propto&
\exp\left\{ -\frac{1}{\sigma^{2}}\,\Phi(|\theta|;\lambda) \right\}.
\label{eq:prior-def}
\end{eqnarray}
This prior is symmetric and unimodal at zero. Since $h(u,\lambda)=0$ for $u\ge a\lambda$,
the penalty $\Phi(r;\lambda)$ becomes constant beyond $r=a\lambda$, so the induced prior
has effectively flat tails on that region. As a result, large coefficients are essentially
unpenalized, and the prior should be interpreted as a nonintegrable
MAP--inducing prior, in the standard penalized--likelihood sense, rather than as a proper
probability distribution.

Combining \eqref{eq:bayes-lik} and \eqref{eq:prior-def}, the negative log posterior (up to constants) is
\begin{eqnarray}
\mathcal{L}(\theta \mid d,\lambda)
&=& \frac{(d-\theta)^{2}}{2\sigma^{2}} + \frac{1}{\sigma^{2}}\,\Phi(|\theta|;\lambda).
\label{eq:neglogpost}
\end{eqnarray}
For $\theta \ne 0$, differentiation gives
\begin{eqnarray}
\frac{\partial\mathcal{L}}{\partial \theta}(\theta \mid d,\lambda)
&=& -\frac{d-\theta}{\sigma^{2}} + \frac{1}{\sigma^{2}}\,\Phi'(|\theta|;\lambda)\,\operatorname{sign}(\theta).
\end{eqnarray}
Since $\Phi'(r;\lambda)=h(r,\lambda)$, stationarity yields
\begin{eqnarray}
d - \theta &=& h(|\theta|,\lambda)\,\operatorname{sign}(\theta).
\label{eq:MAP-eq}
\end{eqnarray}
This is the fixed point equation that characterizes $\rho_{\text{ssc}}$ in the nonzero region: indeed,
 $$
d - \rho_{\text{ssc}}(d,\lambda) = h(d,\lambda)
\quad\text{for }\lambda<|d|<a\lambda,
$$
and for $|d|\ge a\lambda$ the right hand side is zero, giving $\hat{\theta}=d$.

The remaining case $|d|\le\lambda$ corresponds to behavior of the posterior near zero.
On $(0,\lambda]$ one has $h(\theta,\lambda)=\lambda$, so the right derivative of
${\mathcal L}(\theta\mid d,\lambda)$ at $\theta=0$ equals $(-d+\lambda)/\sigma^2$.
Thus $\theta=0$ is a posterior maximizer if and only if $|d|\le\lambda$, and the estimator
is exactly zero on this interval.

This construction links SURE--based tuning of $\lambda$ with a fully specified MAP estimator under a smooth nonconvex prior, rather than treating the threshold as a purely frequentist tuning parameter.

\subsection{Penalty, prior, and posterior}

The penalty $\Phi$ in \eqref{eq:Phi-def} can be written explicitly. After substituting $h$ from \eqref{eq:h-smoothscad-pos}, one obtains
\begin{eqnarray}
\Phi(r;\lambda) &=&
\begin{cases}
\lambda r, & 0 \le r \le \lambda,\\[0.2cm]
\lambda^{2} + \dfrac{(a-1)\lambda^{2}}{\pi}
\left[T + \dfrac{1}{2}\sin(2T)\right], & \lambda < r < a\lambda,\\[0.3cm]
\dfrac{a+1}{2}\lambda^{2}, & r \ge a\lambda,
\end{cases}
\end{eqnarray}
with $T = \dfrac{\pi}{2}\dfrac{r-\lambda}{(a-1)\lambda}$. Inserting this into \eqref{eq:prior-def} gives the smooth SCAD prior density, and combining likelihood and prior gives the unnormalized posterior
\begin{eqnarray}
\pi(\theta \mid d,\lambda,\sigma^{2})
&\propto&
\exp\!\left\{
-\dfrac{1}{2\sigma^{2}}(d-\theta)^{2}
-\dfrac{1}{\sigma^{2}}\,\Phi(|\theta|;\lambda)
\right\}.
\label{eq:ssc-post}
\end{eqnarray}
The posterior mode satisfies \eqref{eq:MAP-eq}. For $|d|$ small, no nonzero solution exists and $\hat{\theta}=0$; once $|d|$ exceeds the model critical value $|d|_{\mathrm{crit}}=\lambda$, a nonzero maximizing branch emerges and moves continuously toward $d$ in the tails.

\subsection{Threshold where the posterior mode leaves zero}

To analyze the transition, focus locally at $\theta=0$. On $(0,\lambda]$, $h(\theta,\lambda)=\lambda$, hence from \eqref{eq:neglogpost}
\begin{eqnarray*}
\frac{\partial}{\partial \theta}\mathcal{L}(\theta \mid d,\lambda)
&=& \frac{\theta-d}{\sigma^{2}} + \frac{\lambda}{\sigma^{2}},
\end{eqnarray*}
and the right derivative at zero is
\begin{eqnarray*}
\left.\frac{\partial}{\partial \theta}\mathcal{L}(\theta \mid d,\lambda)\right|_{\theta\downarrow 0}
&=& \frac{-d+\lambda}{\sigma^{2}}.
\end{eqnarray*}
Thus the posterior decreases immediately to the right of zero when $d<\lambda$, is flat when $d=\lambda$, and increases when $d>\lambda$. A direct comparison of $\mathcal{L}(\theta\mid d,\lambda)$ at $\theta=0$ and $\theta=\varepsilon>0$ shows that as soon as $d>\lambda$, the posterior at a small positive $\theta$ is strictly larger; hence the global maximizer moves off zero at
\begin{eqnarray}
|d|_{\mathrm{crit}} &=& \lambda,
\end{eqnarray}
independently of $a$ and $\sigma^{2}$. For $d>\lambda$ the nonzero mode solves $\theta + h(\theta,\lambda) = d$, and for $d<-\lambda$ the solution is its negative.

\subsection{Extension to the full wavelet model}

In the wavelet model \eqref{eq:normalmodel}, one may place independent priors of the type \eqref{eq:prior-def} on the coefficients $\{\theta_{j,k}\}$. Since both the Gaussian likelihood and the smooth SCAD prior factorize, the negative log posterior separates into univariate terms, and the posterior mode is obtained by applying $\rho_{\text{ssc}}$ coordinatewise. The SURE analysis in Section~\ref{sec:risk-sure-asymptotics} then provides a frequentist risk characterization of this Bayesian estimator, and the asymptotic guarantees of the KS framework apply because the structural and regularity conditions verified in Section~\ref{sec:verification} hold for smooth SCAD.

\medskip
\noindent\textbf{Remark (MAP at zero and subgradient interpretation).}
Because the smooth SCAD prior depends on $|\theta|$, the posterior log density is not differentiable at $\theta=0$, and the stationarity condition must be interpreted in a subgradient sense. In particular, $\theta=0$ is a posterior mode whenever $0$ belongs to the subgradient of $\mathcal{L}(\theta \mid d,\lambda)$ at zero, which occurs exactly when $|d|\le\lambda$. For $|d|>\lambda$ the zero subgradient condition fails, and the posterior admits a nonzero maximizer. This explains why the smooth SCAD estimator is exactly zero on $[-\lambda,\lambda]$ and departs continuously from zero when $|d|$ crosses the critical threshold.


\section{Choice of the threshold for smooth SCAD}
\label{sec:lambda-choice}

We discuss the selection of the main threshold $\lambda$ when the shrinkage rule is smooth SCAD with parameter $a>1$, and relate the optimal $\lambda$ to the signal length $N$, the smoothness exponent $\gamma$, and the shape parameter $a$. We then connect this choice with SURE minimization.

Recall that $N = 2^{J}$ and that, under the wavelet model \eqref{eq:normalmodel},
\begin{eqnarray}
d_{j,k} &=& \theta_{j,k} + \epsilon_{j,k}, \qquad \epsilon_{j,k} \sim N(0,\sigma^{2}),
\end{eqnarray}
with independent $\epsilon_{j,k}$. The underlying function $f$ is assumed to be supported on a finite interval and uniformly Lipschitz regular with exponent $\gamma>0$, so the true coefficients obey a decay bound of the form
\begin{eqnarray}
|\theta_{j,k}| &\le& C_{f}\, 2^{-j(\gamma+1/2)},
\end{eqnarray}
expressing sparsity in the wavelet domain.

For any choice of $h(d,\lambda)$ and the corresponding shrinkage rule $\rho_{h}(d,\lambda)$ defined by \eqref{eq:rho-def}, the mean squared error is given in \eqref{eq:risk}. The threshold $\lambda_{\min}$ that minimizes $R_{J}(\lambda)$ over $\lambda\in[0,\lambda_{U}]$ may be regarded as an oracle choice that depends on the unknown coefficients.

\citet{KudryavtsevShestakov2024} show that for the general class of continuous thresholding rules, and for Lipschitz exponent $\gamma>0$, the oracle risk admits an upper bound as in (\ref{eq:riskrate}),
with $C$ depending on $f$, $\gamma$, and the particular rule, but not on $J$. In the course of their analysis, a balancing of bias and variance terms suggests that, up to lower order corrections, the oracle threshold has the asymptotic scale
\begin{eqnarray}
\lambda_{\mathrm{oracle}}(N,\gamma) &\approx& 
\sigma \sqrt{\frac{4\gamma}{2\gamma+1} \log N}.
\label{eq:lambdaoracle}
\end{eqnarray}
The factor $4\gamma/(2\gamma+1)$ is strictly less than $2$ for every $\gamma>0$, so the oracle threshold is smaller than the universal threshold $\lambda_{U} = \sigma \sqrt{2\log N}$. The dependence on the shape of the rule enters through multiplicative constants in the risk, but not in the leading $\sqrt{\log N}$ scale.

For smooth SCAD, the parameter $a>1$ governs the width of the transition zone $[\lambda,a\lambda]$ where the rule interpolates from strong shrinkage to no shrinkage, see \eqref{eq:h-smoothscad-pos} and \eqref{eq:rho-ssc}. Changing $a$ modifies constants in the risk expression, but the main bias--variance balance remains governed by terms of the form
\begin{eqnarray*}
\lambda^{2} 2^{J/(2\gamma+1)} J^{1/(2\gamma+1)}
\end{eqnarray*}
and
\begin{eqnarray*}
\lambda \exp\Bigl\{-\frac{\lambda^{2}}{2\sigma^{2}}\Bigr\} 2^{J},
\end{eqnarray*}
so that the balancing threshold is of the form \eqref{eq:lambdaoracle}. For fixed $a$ the leading dependence of the best threshold on $N$ and $\gamma$ is unchanged, and $a$ influences only the constant in front of the rate.

In the SURE framework of Section~\ref{sec:risk-sure-asymptotics}, the unbiased risk estimator for smooth SCAD is
\begin{eqnarray}
\widehat{R}_{J}(\lambda) &=& \sum_{j=J-L}^{J-1}\sum_{k=0}^{2^{j}-1} 
F_{\mathrm{ssc}}(d_{j,k},\lambda),
\end{eqnarray}
with $F_{\mathrm{ssc}}$ given by \eqref{eq:Fssc}. The SURE threshold is
\begin{eqnarray}
\lambda_{S} &=& \arg\min_{\lambda \in [0,\lambda_{U}]} \widehat{R}_{J}(\lambda),
\label{eq:lambdaS-smoothscad}
\end{eqnarray}
a specialization of \eqref{eq:lambdaS}. Kudryavtsev and Shestakov prove that for any thresholding function in their admissible class, including smooth SCAD, the SURE threshold $\lambda_{S}$ is with high probability not too small and that $R_{J}(\lambda_{S})$ is asymptotically of the same order as $R_{J}(\lambda_{\min})$.

These results suggest the following picture. For fixed $a>1$ and $\gamma>0$, the risk-optimal threshold (from an oracle standpoint) is on the $\sigma\sqrt{\log N}$ scale, namely
\begin{eqnarray}
\lambda_{\mathrm{oracle}}(N,\gamma,a) &\asymp& 
\sigma \sqrt{\frac{4\gamma}{2\gamma+1} \log N},
\end{eqnarray}
while the universal threshold corresponds to replacing $4\gamma/(2\gamma+1)$ by $2$. The SURE threshold $\lambda_{S}$ is a data--driven choice in $[0,\lambda_{U}]$ with risk close to the minimal risk over that interval, and as $N$ increases, $\lambda_{S}$ behaves like a random perturbation of an oracle threshold of the form \eqref{eq:lambdaoracle}.

\subsection{Approximate choice of $\lambda$ for smooth SCAD}

For smooth SCAD shrinkage it is useful to have a simple analytic suggestion for $\lambda$ when the signal contains $N$ wavelet coefficients. While the exact minimizer of $\widehat{R}_{J}(\lambda)$ is easy to compute on a grid, an explicit approximation provides intuition and a default.

 The classical universal threshold $\lambda_U$ is conservative but guaranteed to remove pure noise. A sparse oracle argument suggests that for many nearly unbiased thresholding rules the optimal threshold is approximated by
\begin{eqnarray}
\lambda_{\mathrm{oracle}}
&\approx& 
\sigma\sqrt{2\log\!\left(\frac{N}{K}\right)},
\end{eqnarray}
where $K$ is the number of nonzero coefficients. Since $K$ is unknown, the universal rule replaces $K$ by $1$, which leads to $\lambda_{U}$. Because smooth SCAD reduces tail bias relative to soft thresholding, its risk optimal threshold is smaller than $\lambda_{U}$.

A convenient empirical heuristic rule is
\begin{eqnarray}
\lambda_{\mathrm{SCAD}}
&\approx&
\sigma \, c(a) \sqrt{2\log N},
\qquad 
c(a)=\sqrt{\frac{a-2}{a-1}},
\label{eq:lambdascadApprox}
\end{eqnarray}
with $a>2$ the SCAD shape parameter. The factor $c(a)$ reflects how sharply the estimator relaxes penalization once $|d|$ exceeds $\lambda$. When $a$ is large, the estimator approaches soft thresholding and $c(a)\to 1$; when $a$ is close to $2$, the shrinkage approaches hard thresholding and $c(a)$ becomes small.
%
%
For the commonly used value $a=3.7$, \eqref{eq:lambdascadApprox} yields
\begin{eqnarray}
\lambda_{\mathrm{SCAD}} &\approx& 0.8\,\sigma\sqrt{2\log N},
\end{eqnarray}
a reduction of about twenty percent relative to the universal threshold.
For smooth SCAD with a raised cosine transition, the added smoothness primarily affects derivative-based quantities (e.g., the SURE expression through $h'$), while the $\sigma\sqrt{2\log N}$ scaling of effective thresholds remains the same as for classical SCAD-type rules.
 In practice the approximation \eqref{eq:lambdascadApprox} provides a reasonable initial choice over a moderate range of $a$.

In summary, smooth SCAD shrinkage suggests a threshold of the form
\begin{eqnarray}
\lambda &\approx& \sigma\sqrt{2\log N}\,\sqrt{\frac{a-2}{a-1}},
\label{eq:lambdascadFinal}
\end{eqnarray}
smaller than the universal threshold yet stable due to the reduced tail bias. This gives a simple and practical rule for selecting $\lambda$ and is consistent with the oracle scale \eqref{eq:lambdaoracle}. In applications, one may combine \eqref{eq:lambdascadFinal} as a starting value with SURE minimization.

\subsection{Choice of the shape parameter $a$}

The smooth SCAD rule also contains the shape parameter $a>1$ that governs how shrinkage transitions from strong attenuation to almost no attenuation. The threshold $\lambda$ controls which coefficients are set to zero, while $a$ controls the width and smoothness of $(\lambda,a\lambda)$ where shrinkage decays. Values of $a$ close to one produce a sharp transition similar to hard thresholding, whereas larger values of $a$ produce a longer and smoother transition.

From the theoretical point of view, $a$ is subject only to mild restrictions. The SURE theory requires that the shrinkage profile $h(d,\lambda,a)$ have bounded first derivatives in both $d$ and $\lambda$ and that $a$ stay in a compact interval bounded away from one. These requirements, verified in Section~\ref{sec:verification}, are satisfied for any fixed $a$ in a range such as $[1.5,4]$. The analysis does not require $a$ to depend on $\sigma^{2}$, $N$, or $\lambda$, in the same spirit as the original SCAD recommendation $a\approx 3.7$ \citep{FanLi2001}.

A natural strategy is to fix $a$ and let SURE determine $\lambda$. Values of $a$ between two and three produce smooth and stable shrinkage; simulations (not reported here) indicate that performance is not highly sensitive to the precise choice. Once $a$ is fixed, SURE selection proceeds as in \eqref{eq:lambdaS-smoothscad}.

If one prefers a mild dependence on $N$ and $\sigma$, one can align the end of the transition zone with the universal threshold. Since $\lambda_{U}=\sigma \sqrt{2\log N}$, requiring $a\lambda\approx\lambda_{U}$ leads to
\begin{eqnarray}
a &\approx& \frac{\sigma \sqrt{2\log N}}{\lambda}.
\end{eqnarray}
With $\lambda$ chosen by SURE, this provides a data--driven rule for $a$ subject to bounds such as $a\in[1.5,4]$. A more elaborate alternative is to select $a$ jointly with $\lambda$ by minimizing SURE over a small grid $a\in\{2,2.5,3,3.5\}$; for each $a$ one finds $\lambda_{S}(a)$ and chooses the pair with smallest SURE risk.

To summarize, the choice of $a$ is less delicate than the choice of $\lambda$. The main asymptotic dependence on $N$ and $\sigma$ is carried by $\lambda$, while $a$ tunes the smoothness of the transition. A fixed moderate $a$ combined with SURE selection of $\lambda$ already yields an effective procedure.

\subsection{Adaptive smooth SCAD with
level-dependent thresholds}

So far the parameters $\lambda$ and $a$ were global. In many situations the variability of the empirical wavelet coefficients changes substantially with level $j$. It is often advantageous to allow level-dependent thresholds. As before, we consider detail wavelet coefficients from decomposition levels $j = J-L, \dots, J-1$, where $J-1$ corresponds to the finest scale of detail coefficients and $J-L$ to the coarsest one included in shrinkage.

To incorporate heterogeneity over $a$ and $\lambda$, we introduce a level-dependent version of smooth SCAD. For each wavelet coefficient $d$ from level $j$, we choose a main threshold $\lambda_{j}$ and a transition parameter $a_{j} > 1$, and define
\begin{eqnarray}
h_{j}(d,\lambda_{j}) &=&
\begin{cases}
\lambda_{j}\,\operatorname{sign}(d), & 0 < |d| \le \lambda_{j},\\[0.2cm]
\lambda_{j}\cos^{2}\!\left(\dfrac{\pi}{2}\,\dfrac{|d|-\lambda_{j}}{(a_{j}-1)\lambda_{j}}\right)\operatorname{sign}(d), 
& \lambda_{j} < |d| < a_{j}\lambda_{j},\\[0.3cm]
0, & |d| \ge a_{j}\lambda_{j},
\end{cases}
\end{eqnarray}
and the adaptive smooth SCAD rule
\begin{eqnarray}
\rho_{\mathrm{ssc},j}(d,\lambda_{j}) &=&
\begin{cases}
0, & |d| \le \lambda_{j},\\[0.1cm]
d - h_{j}(d,\lambda_{j}), & \lambda_{j} < |d| < a_{j}\lambda_{j},\\[0.1cm]
d, & |d| \ge a_{j}\lambda_{j}.
\end{cases}
\end{eqnarray}
The shrinkage applied to $d_{j,k}$ is then $\rho_{\mathrm{ssc},j}(d_{j,k},\lambda_{j})$.

This construction preserves oddness, continuity, and bounded derivative properties at each level, since each triple $(h_{j},\lambda_{j},a_{j})$ satisfies the same structural assumptions as the global version. All SURE calculations extend levelwise. At level $j$ the unbiased risk estimate is
\begin{eqnarray}
\widehat{R}_{j}(\lambda_{j}) &=& \sum_{k=0}^{2^{j}-1} F_{\mathrm{ssc}}(d_{j,k},\lambda_{j}),
\end{eqnarray}
with $F_{\mathrm{ssc}}$ as in \eqref{eq:Fssc}, but with $h$ and its derivative replaced by $h_{j}$ and its derivative. The global risk estimate is
\begin{eqnarray}
\widehat{R}_{J}(\lambda_{J-L},\dots,\lambda_{J-1}) &=& \sum_{j=J-L}^{J-1} \widehat{R}_{j}(\lambda_{j}).
\end{eqnarray}

For a fixed transition parameter $a$, a simple adaptive scheme is to let SURE pick $\lambda_{j}$ independently at each level as
\begin{eqnarray}
\lambda_{j,S} &=& \arg\min_{\lambda_{j}\in[0,\lambda_{U}]} \widehat{R}_{j}(\lambda_{j}).
\end{eqnarray}
This extension of SureShrink keeps each level adaptive to its own coefficient structure. The admissibility conditions remain satisfied because the KS analysis controls each level separately and the sums of SURE terms remain well behaved.

To improve stability, especially at coarser levels where the number of coefficients $2^{j}$ is smaller and the SURE surface is more variable, it is natural to impose a lower floor on the selected threshold. A practical choice is
\begin{eqnarray}
\lambda_{j}^{\star}
&=&
\min\Bigl\{\lambda_{U},\ \max\{\lambda_{j,S},\ c\,\sigma\}\Bigr\},
\end{eqnarray}
where $c$ is a fixed constant. Values in the range $c\in[2,3]$ are reasonable in practice, providing sufficient protection against excessive noise retention while allowing the SURE-selected threshold to dominate when it is already conservative. Without such a floor, $\lambda_{j,S}$ may occasionally be too small, leading to residual noise at levels with few coefficients. The floor therefore represents a simple and effective stabilization of the levelwise SURE rule; more elaborate constructions are possible by incorporating information about the smoothness of the underlying function.

A second scheme is to let both $\lambda_{j}$ and $a_{j}$ vary with $j$. Larger $a_{j}$ widen the transition region $[\lambda_{j},a_{j}\lambda_{j}]$ and produce gentler shrinkage at coarse scales, where coefficients tend to contain genuine signal; smaller $a_{j}$ make the rule approach hard thresholding more rapidly, which may be appropriate for the finest, noise–dominated levels. One can fix a small set of candidate values for $a_{j}$, such as $a_{j}\in\{1.5,2,3\}$, and choose $a_{j}$ by SURE on a small grid, or impose a deterministic levelwise schedule, for example taking smaller $a_{j}$ at fine scales and larger $a_{j}$ at coarse scales. As long as $a_{j}>1$ and uniformly bounded, the structural assumptions continue to hold and the SURE expressions retain their form.


\section{Numerical illustration}
\label{sec:numerical}

To illustrate the practical behavior of the smooth SCAD denoiser, we conducted a controlled Monte Carlo study using the four canonical Donoho-Johnstone test functions: Doppler, Blocks, HeaviSine, and Bumps. These functions are widely adopted benchmark signals in the wavelet–denoising literature \citep{DonohoJohnstone1994, DonohoJohnstone1995}, each representing a different mode of nonparametric difficulty: Doppler exhibits spatially varying oscillations; Blocks contains sharp discontinuities; HeaviSine combines smooth and abrupt features; and Bumps contains multiple heterogeneous local spikes.

For each signal, independent Gaussian white noise was added at controlled SNRs. Following the statistical convention, SNR is defined as the ratio of signal variance to noise variance; we fix $\mathrm{Var}(\epsilon)=1$ and rescale each clean signal so that $\mathrm{Var}(f)/\mathrm{Var}(\epsilon)=\mathrm{SNR}$. Experiments were performed for $\mathrm{SNR}=3,5,7$ and $10$, and for sample sizes $N=512, 1024$ and $N=2048$, though only representative results are shown Table \ref{table:dj10247} for brevity. In all cases, $M=1000$ independent Monte Carlo replications were generated, and results are summarized through the average mean–squared error (AMSE) and its Monte Carlo standard deviation.

Wavelet transforms were computed using orthonormal compactly supported filters standard in the denoising literature: Symmlet 4 for Doppler and HeaviSine, Haar for Blocks (to provide an unbiased representation of discontinuities), and Daubechies 3 for Bumps. 
Denoising was performed coefficientwise using four shrinkage rules: universal hard
thresholding, universal soft thresholding, classical SCAD \citep{FanLi2001} with oracle
selection of $\lambda$, and its smooth raised cosine analogue (smooth SCAD). The oracle
tuning of SCAD and smooth SCAD is used here to isolate the effect of the shrinkage shape;
universal hard and soft thresholds serve as fixed plug--in baselines.

\begin{figure}[t]
\centering

\begin{minipage}{0.48\textwidth}
\centering
\includegraphics[width=\linewidth]{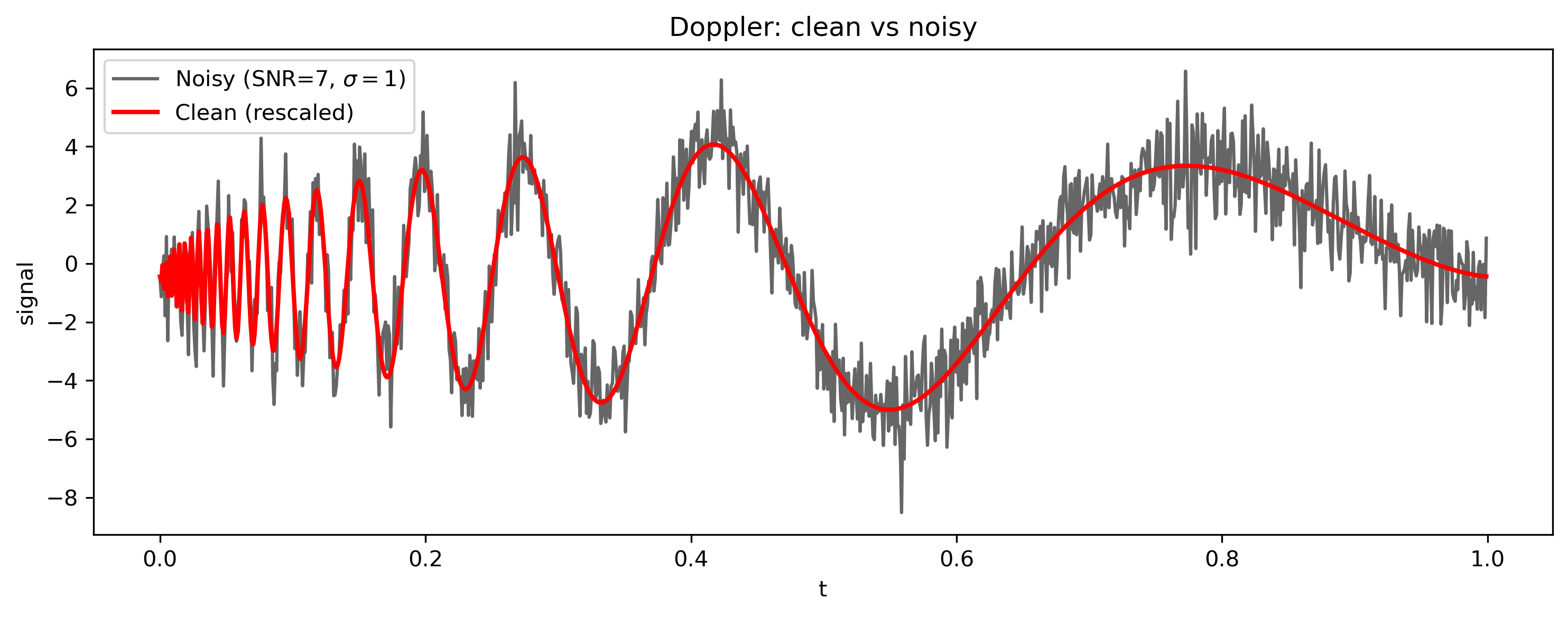}
\\[-2mm]
{\small (a)}
\end{minipage}
\hfill
\begin{minipage}{0.48\textwidth}
\centering
\includegraphics[width=\linewidth]{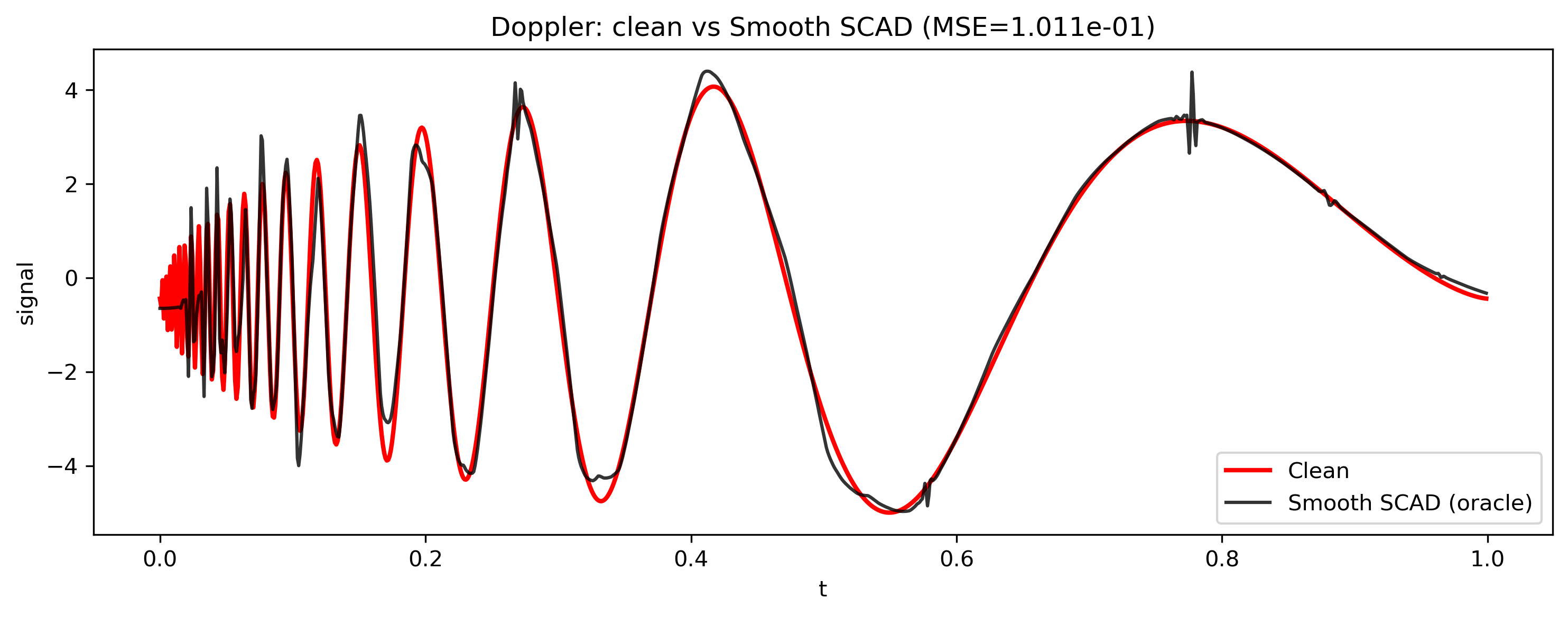}
\\[-2mm]
{\small (b)}
\end{minipage}
\caption{\small (a) Doppler test signal of length $N=1024$ with additive Gaussian noise, rescaled to achieve $\mathrm{SNR}=7$ (variance ratio);
(b) Smooth SCAD reconstruction using oracle–selected threshold $\lambda^* = \arg \min_\lambda \mbox{MSE}(\lambda) \approx 2.3485,$  showing effective recovery of structure and suppression of high-frequency artifacts. The oracle–selected $\lambda^*$ is smaller than the universal threshold, which in this case is $\lambda_{U} = \sqrt{2 \log(1024)} = 3.724$.}
\label{fig:smoothscad}
\end{figure}

Figure \ref{fig:smoothscad} displays a Doppler data set corrupted by Gaussian noise at $\mathrm{SNR}=7$, together with its denoised reconstruction obtained using the Smooth SCAD shrinkage rule.

Across all configurations, smooth SCAD consistently improved upon SCAD, and both substantially outperformed universal thresholding, particularly in high-frequency or discontinuous settings. The gains in average MSE (over $M=1000$ runs) were most pronounced for Blocks and Bumps, where noise suppression without excessive bias is essential. Table \ref{table:dj10247} 
reports representative results for $N=1024$, $\mathrm{SNR}=7,$ and $\sigma=1$. 

\begin{table}[htb]
\centering
\caption{\small AMSE and standard deviation of MSE over $M=1000$ runs, $N=1024$ and SNR = 7.0.}
\label{tab:amse_std_four_signals}
\medskip
\begin{tabular}{lcccc}
\hline
Signal & Hard & Soft & SCAD & Smooth SCAD \\
\hline
Doppler & 1.5513e-01 & 4.0264e-01 & 1.3505e-01 & 1.3188e-01 \\
 & (1.8200e-02) & (3.2646e-02) & (1.4472e-02) & (1.4609e-02) \\
\hline
Blocks & 2.0078e-01 & 6.6805e-01 & 1.5286e-01 & 1.4484e-01 \\
 & (2.7987e-02) & (4.4013e-02) & (1.8624e-02) & (1.9502e-02) \\
\hline
HeaviSine & 7.7260e-02 & 1.0477e-01 & 6.8234e-02 & 6.5714e-02 \\
 & (1.5162e-02) & (1.1461e-02) & (1.0011e-02) & (1.0515e-02) \\
\hline
Bumps & 2.9867e-01 & 9.5999e-01 & 2.3165e-01 & 2.2799e-01 \\
 & (3.2696e-02) & (5.2262e-02) & (2.0762e-02) & (2.1343e-02) \\
\hline
\end{tabular}
\label{table:dj10247}
\end{table}

 The simulation engine was implemented in Python  and is fully reproducible; Jupyter notebooks can be found
at \url{https://github.com/BraniV/SmoothSCAD}.


\section{Conclusions}
\label{sec:conclusions}

The smooth SCAD rule developed here refines the classical SCAD threshold by replacing its piecewise linear transition with a raised cosine. This modification preserves the essential SCAD behavior while producing a shrinkage profile that is continuous on $\mathbb{R}$ and continuously differentiable away from the main threshold. In doing so, it places the rule entirely inside the continuous thresholding class for which Stein's unbiased risk estimate applies.

Because the smooth SCAD shrinkage profile satisfies the structural and regularity conditions required by the SURE framework, unbiased computation of the risk and its minimizer is available without further adjustment. The asymptotic conclusions established for continuous thresholding rules---oracle order convergence of the minimal risk, asymptotic normality of the centered SURE risk, and strong consistency under normalization---carry over directly. These properties hold levelwise and extend to level--dependent and adaptive constructions.

The Bayesian interpretation further clarifies the method. The raised cosine transition induces a smooth nonconvex penalty whose posterior mode reproduces the smooth SCAD estimator exactly, linking penalized likelihood arguments with wavelet shrinkage while also highlighting the role of the transition width in mediating sparsity and tail behavior. The raised cosine transition is one convenient smooth choice; other smooth interpolants
(e.g., spline--based transitions) could satisfy the same SURE admissibility conditions while
altering higher order derivative behavior, and may be worth exploring.

In practice, smoothness yields computational advantages. SURE minimization over $\lambda$ is numerically stable, and shrinkage of borderline coefficients is more regular than under piecewise linear SCAD. A simple analytic expression for the threshold scale provides a reasonable starting value, while the SURE principle supplies data adaptivity across scales.

Several directions follow naturally. First, the raised cosine construction may be used to obtain smooth analogues of other nonconvex penalties, including MCP and smoothly clipped $\ell_{0}$–type rules, which would likewise fall inside the continuous SURE class. Second, smooth SCAD can be embedded into iterative thresholding and proximal–gradient schemes for linear inverse problems, where differentiability of the shrinkage operator facilitates convergence analysis. Third, extensions to non–Gaussian noise models are possible by replacing Stein identities with their exponential–family or Tweedie–type analogues, potentially yielding SURE variants for heavy–tailed or heteroscedastic data.

To our knowledge, smooth SCAD is the first SCAD--type wavelet shrinkage rule that simultaneously satisfies the continuous SURE conditions of the KS framework and admits an explicit nonconvex prior whose posterior mode coincides with the estimator. The result is a transparent, flexible, and practical shrinkage rule that fits naturally into wavelet–based statistical methodology.

\bibliography{references}

\end{document}